\documentclass[aps,prb,twocolumn,superscriptaddress,showpacs,floatfix,longbibliography]{revtex4-2}
\usepackage{amsmath,amssymb,amsfonts,float,graphics,epsfig,epstopdf,color,verbatim,tabularx,bm,multirow,appendix,hyperref}
\usepackage{lmodern}
\usepackage{ytableau}
\usepackage{color}
\usepackage{bm}

\newcommand{\beq} {\begin{equation}}
	\newcommand{\eeq} {\end{equation}}
\newcommand{\bea} {\begin{eqnarray}}
	\newcommand{\eea} {\end{eqnarray}}
\newcommand{\be} {\begin{equation}}
	\newcommand{\ee} {\end{equation}}

\DeclareMathOperator{\Tr}{Tr}

\def\bk{{\mathbf{k}}}
\def\bp{{\mathbf{p}}}

\def\bq{{\mathbf{q}}}

\def\bG{{\mathbf{G}}}

\begin{document}
	
\title {Superconductivity and bosonic fluid emerging from moir\'e flat bands}

\author{Xu Zhang}
\affiliation{Department of Physics and HKU-UCAS Joint Institute of Theoretical and Computational Physics, The University of Hong Kong, Pokfulam Road, Hong Kong SAR, China}
\author{Kai Sun}
\email{sunkai@umich.edu}
\affiliation{Department of Physics, University of Michigan, Ann Arbor, Michigan 48109, USA}
\author{Heqiu Li}
\affiliation{Department of Physics, University of Michigan, Ann Arbor, Michigan 48109, USA}
\affiliation{Department of Physics, University of Toronto, Toronto, Ontario M5S 1A7, Canada}
\author{Gaopei Pan}
\affiliation{Beijing National Laboratory for Condensed Matter Physics and Institute of Physics, Chinese Academy of Sciences, Beijing 100190, China}
\affiliation{School of Physical Sciences, University of Chinese Academy of Sciences, Beijing 100190, China}
\author{Zi Yang Meng}
\email{zymeng@hku.hk}
\affiliation{Department of Physics and HKU-UCAS Joint Institute of Theoretical and Computational Physics, The University of Hong Kong, Pokfulam Road, Hong Kong SAR, China}

\date{\today}
		
\begin{abstract}
With the evidence of inter-valley attraction-mediated by phonon or topological fluctuations, we assume the inter-valley attraction and aim at identifying universal properties of moir\'e flat bands that shall emerge. We show that by matching the interaction strength of inter-valley attraction with  intra-valley repulsion, the flat-band limit becomes exactly solvable. Away from the flat-band limit, 
the system can be simulated via quantum Monte Carlo (QMC)  methods without sign problem for any fillings. Combining analytic solutions with large-scale numerical simulations, we show that 
upon increasing temperature, the superconducting phase melts into a bosonic fluid of Cooper pairs with large/diverging compressibility. 
In contrast to flat-band attractive Hubbard models, where similar effects arise only for on-site interactions, our study indicates this physics is a universal property of moir\'e flat bands, regardless of microscopic details such as the range of interactions and/or spin-orbit couplings. 
At higher temperature, the boson fluid phase gives its way to a pseudo gap phase, where some Cooper pairs  are torn apart by thermal fluctuations, resulting in fermion density of states  inside the gap. Unlike the superconducting transition temperature, which is very sensitive to doping and twisting angles, the gap and the temperature scale of the boson fluid phase and the pseudo gap phase are found to be nearly independent of doping level and/or flat-band bandwidth. The relevance of these phases with experimental discoveries in the flat band quantum moir\'e materials is discussed. 
\end{abstract}

\maketitle

\section{Introduction}
As one of the most intriguing development in 2D materials, moir\'e superlattices offer a new opportunity to access novel quantum states and quantum phenomena~\cite{Novoselov2016,andrei2020graphene,dean2010boron,wang2013one}, such as flat bands at magic angle twisted bilayer graphene (TBG) or  transition metal dichalcogenides (TMD)~\cite{Trambly2010,Bistritzer_TBG,MacDonald2019}.
Recently, these systems were brought to the forefront of research by a series of intriguing experimental discoveries, such as correlated insulating states, continuous Mott transition and superconductivity~\cite{cao2018unconventional,cao2018correlated,wang2020correlated,doi:10.1126/science.aav1910,doi:10.1126/science.aaw3780,lu2019superconductors,lu2019superconductors,doi:10.1126/science.aay5533,Zondiner2020,saito2020independent,saito2020independent,stepanov2020untying,PhysRevLett.124.076801,polshyn2019large,xie2019spectroscopic,jiang2019charge,choi2019electronic,wong2020cascade,liu2020tunable,cao2020tunable,shen2020correlated,PhysRevLett.123.197702,chen2019evidence,chen2019signatures,chen2020tunable,zhou2021superconductivity,NWang2020,Ghiotto2021,TingxinLi2021}. 
In addition to TBG, superconductivity has also been observed in other 2D materials such as MoS$_2$~\cite{JMLu2015,saito2016superconductivity}, NbSe$_2$~\cite{XiaoxiangXi2016} and possibly in twisted bilayer and double-bilayer WSe$_2$~\cite{wang2020correlated,NWang2020}, which is believed to be due to inter-valley attractions~\cite{phononmos21,phononmos22,mos2scsymmetry}. In TBG, inter-valley attractions has also been considered as a key candidate mechanism for the superconducting state, though the origin of such attractions is still under investigation, i.e. whether it is phononic or has some more exotic (and even topological) mechanism~\cite{BABphonon,gonzalez2019kohn,lee2019theory,ferroSU4,chatterjee2020skyrmion,skyrmionsc,PhysRevX.8.031089,PhysRevLett.121.087001,PhysRevB.99.121407,PhysRevB.98.075154,PhysRevLett.121.257001,PhysRevLett.121.257001,PhysRevX.8.041041,you2019superconductivity,PhysRevB.98.241407,guinea2018electrostatic,PhysRevB.98.085436,PhysRevB.101.165141,lewandowski2021does}. 

Here, we focus on universal principles/properties that flat-band moir\'e superconductors shall obey/exhibit, in that, we introduce an inter-valley attractive interaction and study its nontrivial consequence. By matching inter-valley attraction strength with intra-valley repulsion,  we find that at the flat band limit, such systems can be solved exactly. Away from the flat-band limit,  although an exact solution is absent, the model can be simulated via the momentum-space quantum Monte Carlo (QMC) method~\cite{XuZhang2021,JYLee2021,GaopeiPanValley2021} without suffering the sign problem at any fillings, as the case in attractive Hubbard model ~\cite{PhysRevB.28.4059,PhysRevLett.69.2001,PhysRevLett.75.312,PhysRevB.54.1286,PhysRevB.69.184501}. By combining exact analytic solution, exact diagonalization at small sizes and the fully momentum-space QMC, we show that moir\'e flat-band superconductors exhibit a rich phase diagram, in sharply contrast to conventional BCS superconductors. As shown in Fig.~\ref{fig:fig2}(a-b), at the temperature above the superconducting dome, the system doesn't directly transform into a Fermi liquid. Instead, it first turns into a super-compressible bosonic fluid phase, where fermionic excitations are fully gapped but the compressibility is high and increases/diverges upon cooling.
This physics is in strong analogy to the flat-band attractive Hubbard model~\cite{PhysRevB.94.245149, PhysRevB.102.201112}, but also with clear differences. In the flat-band Hubbard model, the same type of physics only arises when the interactions are strictly on-site, and non-onsite interactions, such as nearest-neighbor, quickly leads to other physics, e.g., phase separation~\cite{PhysRevB.102.201112}. In contrast, for moire flat bands, our studies indicate that the exactly solution and related phenomena are extremely robust and fully insensitive to such microscopic details. No matter the interaction is short-range or longe-range and no matter spin-orbit coupling is weak (e.g. graphene) or strong (TMD), our exact solution and all qualitative features remain the same. This robust is of crucial important for experimental study of moir\'e lattices, where on-site interaction are not expected to be dominant and spin-orbit effect may or may not be strong depending on the materials. As one further increases the temperature, this bosonic fluid gives its way to a pseudogap phase, where fermion states start to emerge inside the gap and gradually fill it up upon increasing the temperature.  These nontrivial sequence of phase transitions/crossovers arise at filling range larger than superconducting dome and survive to temperature much higher than the superconducting transition $T_c$.

The bosonic fluid phase can be viewed as a liquid of Cooper pairs, where although the Cooper pairs have fully formed, (quasi)-long-range phase coherence has not yet been developed due to strong fluctuations. Experimentally, key signature of this phase is a large single-particle gap around the Fermi energy, combined with a high/diverging compressibility.
The pseudogap phase is a partially melted boson liquid, where thermal fluctuations start to tear apart some Cooper pairs , feeding single fermion spectral weight into the energy gap.

Simulations and analytic theory also indicate that the crossover temperatures between normal fluid and the pseudogap phase (or between the pseudogap and the bosonic fluid phases)
are dictated by the energy scale of the inter-valley attractions, and is nearly independent of filling levels or flat-band band width. In contrast, the superconducting transition temperature depends strongly on the bandwidth of the flat band, as well as filling fractions. In the QMC simulations, a superconducting dome is observed, qualitatively consistent with the experimental observations in TBG and TMD systems with chemical potential partially filling all flat bands~\cite{lu2019superconductors,Zondiner2020,wang2020correlated,NWang2020}. As the band width reduces to zero (i.e. towards the flat band limit), the height of the dome (i.e. superconducting transition temperature) decreases to zero, although the non-interacting density of states (DOS) for the flat band diverges. 

\section{Model}
We consider a generic system with two valleys, labeled by the valley index $\tau$ and $-\tau$ respectively, connected by the time-reversal  transformation ($T \tau = -\tau$). 
In the flat-band limit, kinetic energy can be dropped once other bands are projected out. 
For interactions, we define it in the momentum space as 
\begin{equation}
	H_{I}=\frac{1}{2 \Omega} \sum_{\mathbf{G}} \sum_{\mathbf{q} \in mBZ} V(\mathbf{q+G}) \delta \rho_{\mathbf{q+G}} \delta \rho_{\mathbf{-q-G}},
\label{eq:eq1}
\end{equation}
where $\mathbf{q}$ is the momentum transfer in a moir\'e Brillouin zone (mBZ) and $\mathbf{G}$ is the moir\'e reciprocal lattice vector. We set the interaction strength $V(\mathbf{q+G})$ to be an arbitrary positive function and
$\delta \rho_{\mathbf{q+G}}$ is the density difference between two valleys
\begin{equation}
\delta \rho_{\mathbf{q+G}}=\rho_{\tau;\mathbf{q+G}}-\rho_{-\tau;\mathbf{q+G}}.
\label{eq:delta_rho}
\end{equation}
where $\rho_\tau$ and $\rho_{-\tau}$ are the fermion density from the two valleys. 
In band basis, $\delta \rho$ can be projected to the flat band
\begin{eqnarray}
\delta \rho_{\mathbf{q+G}}&=&\sum_{\mathbf{k},m,n} [\lambda_{m,n,\tau}(\mathbf{k}, \mathbf{k+q+G})c_{\mathbf{k}, m, \tau}^{\dagger} c_{\mathbf{k+q}, n, \tau} \nonumber\\
&&-\lambda_{m,n,-\tau}(\mathbf{k}, \mathbf{k+q+G})c_{\mathbf{k},m,-\tau}^{\dagger} c_{\mathbf{k+q},n,-\tau}]
\end{eqnarray}
where the form factor $\lambda$ is computed via the unitary transformation between the plane-wave basis and the band basis (see Supplemental Material (SM)~\ref{app:app1} for details). 
This interaction contains inter-valley attraction and intra-valley repulsion with the same strength $V$, and this is how inter-valley attractions are introduced in our model. With this interaction, this model is exactly solvable in the flat-band limit, while away from the exactly-solvable limit (i.e. with finite band width and chemical potential), it can be simulated via QMC without sign problem.

\section{Exact Solution}
After dropping the trivial kinetic energy, the flat-band limit of our Hamiltonian is reduced
to $H=H_I$ [Eq.~\eqref{eq:eq1}], which can be solved exactly, due to an emergent SU(2) symmetry with generators
\begin{equation}
\sigma_x=\Delta+\Delta^\dagger, \sigma_y=i (\Delta-\Delta^\dagger), \sigma_z=\hat{N}_p-N_d,
\end{equation}
where $\Delta^{\dagger} = \sum_{\mathbf{k},m} c_{\mathbf{k},m,\tau}^{\dagger} c_{\mathbf{-k},m,-\tau}^{\dagger}$ and $\Delta = \sum_{\mathbf{k},m} c_{\mathbf{-k},m,-\tau} c_{\mathbf{k},m,\tau}$ creates/annihilates one inter-valley Cooper pair and $\hat{N}_p$ is the particle number operator of flat bands. The constant $N_d$ is the max electron number that these flat bands can host in one valley. It is easy to verify that these three operators obey the su(2) algebra $[\sigma_i,\sigma_j]=2i \epsilon_{ijk}\sigma_k$ and they all commute with $\delta\rho$ and $H_I$, $[\sigma_i,H_I]=0$. In other words, these three operators generate a SU(2) symmetry group.

\begin{figure}[htp!]
	\includegraphics[width=1\columnwidth]{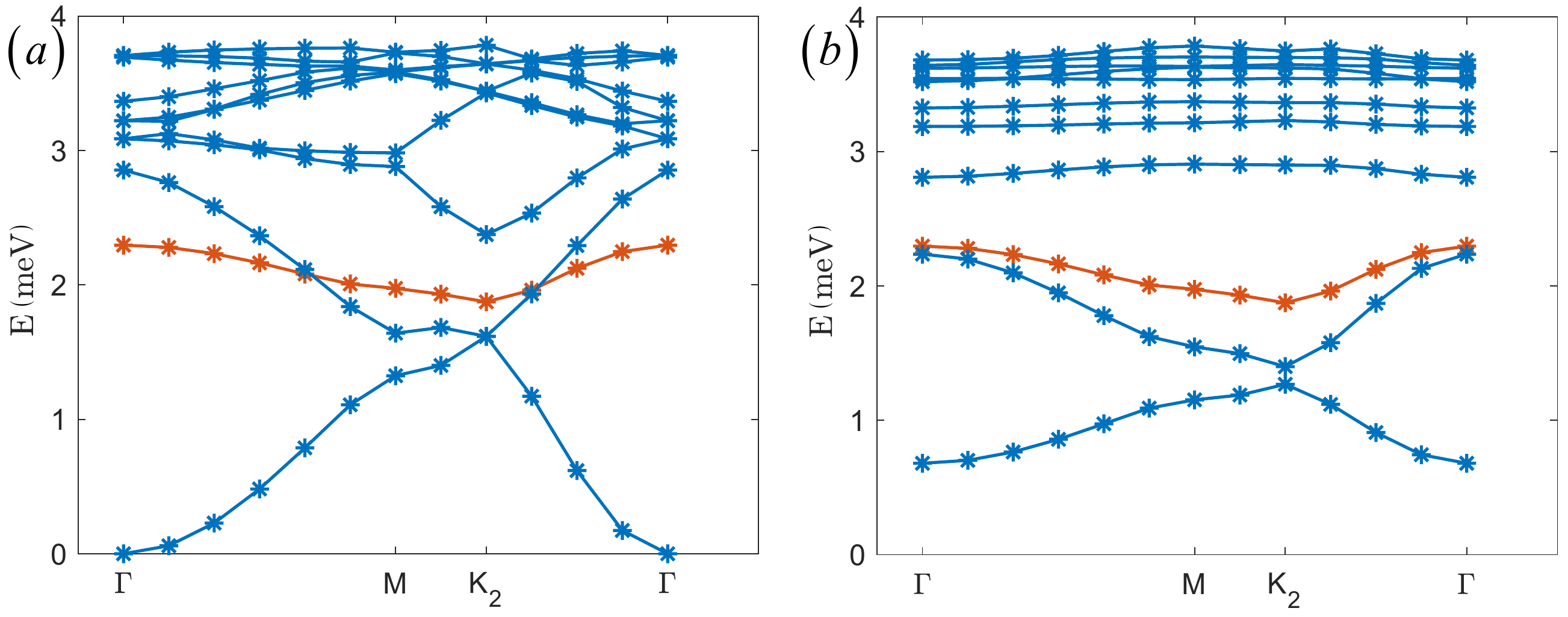}
	\caption{Excitation spectrum for inter-valley attractive and repulsive model where the system size is $12\times12$ and parameters come from~\ref{app:app1}. (a) Single-paritcle excitation (red line) and Goldstone excitation (blue lines) along high symmetry line for inter-valley attractive model. (b) Single-paritcle excitation (red line) and gapped charge neutral excitation (blue lines) along high symmetry line for inter-valley repulsive model.}
	\label{fig:fig1}
\end{figure}

This emergent SU(2) symmetry and exact solution are in analogy to the flat-band Hubbard model~\cite{PhysRevB.94.245149} and the SU(4) emergent symmetry of TBG flat bands ~\cite{bernevig2020tbg3,PhysRevX.10.031034,PhysRevLett.125.257602,PhysRevLett.122.246401}, but there are some key differences. For the Hubbard model, the emergent symmetry and exact solution only arises when the interaction is on-site, and interactions beyond on-site (e.g., nearest-neighbor) takes away the exact solution and results in other instability like phase separation~\cite{PhysRevB.102.201112}. In contrast, our exact solution is insensitive to the range and/or the functional form of interactions. It is also worthwhile to highlight that the attractive Hubbard model can be exactly mapped to a repulsive Hubbard at half filling via a particle-hole transformation. Such a mapping doesn't exist in general for moir\'e flat bands, because the particle-hole transformation will change the $\lambda$ function used in the flat-band project. For inter-valley repulsive model at half filling,
\begin{eqnarray}
	\delta\rho_{\mathbf{q+G}}^{\prime}&&=\rho_{\tau;\mathbf{q+G}}+\rho_{-\tau;\mathbf{q+G}} \nonumber\\
	&&=\sum_{\mathbf{k},m,n} \lambda_{m,n,\tau}(\mathbf{k}, \mathbf{k+q+G}) \nonumber\\
	&&\times(c_{\mathbf{k}, m, \tau}^{\dagger} c_{\mathbf{k+q}, n, \tau} -\bar{c}_{\mathbf{k},m,-\tau}^{\dagger}\bar{c}_{\mathbf{k+q},n,-\tau})
\end{eqnarray}
Here, we use particle-hole transformation $\bar{c}_{\mathbf{k},m,-\tau}=c_{\mathbf{-k},m,-\tau}^{\dagger}$. It is obvious that ground states for a general inter-valley repulsive model have valley polarized $Z_2$, not SU(2) symmetry. This difference can also be seen in charge neutral excitation spectrums as shown in Fig.~\ref{fig:fig1}. One can see although the two models share the same single-particle excitation spectrum as the red lines in Fig.~\ref{fig:fig1}(a,b), their two-particle spectra are totally distinct (i.e. continuous excitation in Fig.~\ref{fig:fig1}(a) and gapped in Fig.~\ref{fig:fig1}(b) as the blue lines indicate). The attractive model is gapless due to the Goldstone mode from the SU(2) symmetry, while the repulsive one is gapped due to the absence of SU(2) symmetry and Goldstone modes. We will show how to derive exact solutions briefly below and leave details in~\ref{app:app4}.

Because our Hamiltonian is semi-positive definite [$V>0$ in Eq.~\eqref{eq:eq1}], two obvious zero-energy ground states can be immediately identified: the empty and fully filled states labeled by $\left| \psi_{0}\right\rangle $ and $\left| \psi_{2N_d}\right\rangle $, where $\delta \rho \left| \psi_{0}\right\rangle=\delta \rho \left| \psi_{2N_d}\right\rangle=0$.
In terms of the SU(2) symmetry group, these two states are fully polarized states of  the $\sigma_z$ operator, known as the highest weight states, where the eigenvalues of $\sigma_z$ reach the highest/lowest possible values $\pm N_d$. 
Due to the SU(2) symmetry, any SU(2) rotation of these two ground states must also be a degenerate ground state. Here, we can use $\Delta^\dagger$ and $\Delta$ as raising and lowering operators of the su(2) algebra, and generate all the degenerate ground states from the empty state $\left| 0\right\rangle \equiv\left| \psi_{0}\right\rangle$
\begin{align}
\left| \psi_{2n}\right\rangle \equiv \sqrt{\frac{(N_d-n)!}{n! (N_d)!}}(\Delta^{\dagger})^n\left| 0\right\rangle
\end{align}
where $0\le n \le N_d$ and $\left| \psi_{2n}\right\rangle$ is the degenerate ground state with $2n$ fermions.
Another way to understand these degenerate ground states is to realize $[H_I,\Delta^\dagger]=0$ implies that it costs no energy to create a Cooper pair. Thus, we can introduce arbitrary number of Cooper pairs to the empty state, and obtain a degenerate ground state $\left| \psi_{2n}\right\rangle$ with $n$ Cooper pairs.
It is worthwhile to highlight that the BCS wavefunction $\psi_{BCS}=\frac{1}{\mathcal{N}}\exp(\frac{v}{u} \Delta^{\dagger})\left| 0\right\rangle$ is also an exact ground state of this model. But this one is nothing special and just one of $N_d+1$ degenerate ground states. The BCS wavefunction may not be favored when kinetic term and chemical potential is introduced, since chemical potential term will favor a certain filling but not mixing states with different particle number.

In the exact solution, single particle correlation function can be computed by noticing $c_{\mathbf{k},\tau}^{\dagger} \left| \psi_{2n}\right\rangle = \sqrt{\frac{N_d-n}{N_d}} \left| \psi_{2n+1,\mathbf{k},\tau}\right\rangle$, $c_{\mathbf{k},\tau} \left| \psi_{2n}\right\rangle = \sqrt{\frac{n}{N_d}} \left| \psi_{2n-1,\mathbf{-k},-\tau}\right\rangle$, where $\left| \psi_{2n+1,\mathbf{k},\tau}\right\rangle\equiv\sqrt{\frac{(N_d-n-1)!}{n!(N_d-1)!}}(\Delta^{\dagger})^n c_{\mathbf{k},\tau}^{\dagger}\left| 0\right\rangle$. As shown by red line in Fig.~\ref{fig:fig1}, single particle excitation is fully gapped and independent with inter-valley interaction (no matter repulsion or attraction). For repulsive interactions, the gap is an insulating gap without other charged excitation inside~\cite{lian2020tbg4,bernevig2020tbg5,PhysRevX.10.031034,PhysRevLett.125.257602,PhysRevLett.122.246401}. In our model, this gap is the Cooper gap, i.e., the energy cost to break a Cooper pair. The Cooper gap scales linearly with interaction energy $V$ and there are continuous charged excitations within the gap as the blue line in Fig.~\ref{fig:fig1}(a). Thus, at low temperature $T\ll V$, all electrons are paired into Cooper pairs, i.e., the system is a fluid of Cooper pairs without unpaired fermions.

We can also compute the correlation function of Cooper pairs $\left\langle \Delta(t) \Delta^\dagger(0)\right\rangle$ by noticing $\Delta^\dagger \left| \psi_{2n}\right\rangle = \sqrt{(N_d-n)(n+1)} \left| \psi_{2n+2}\right\rangle$. Because $[H_I,\Delta]=[H_I,\Delta^\dagger]=0$, this correlation function is time-independent at any temperature. At $T=0$,  this correlation function is
\begin{eqnarray}
\left\langle \Delta(t) \Delta^\dagger(0)\right\rangle &=& \frac{N_d(N_d+2)}{6}\sim \frac{N_d^2}{6},
\label{eq:eq6}
\end{eqnarray}
which is in good agreement with QMC simulations (see SM Fig.~\ref{fig:fig4}(a,b)). It is also worthwhile to point out that in the thermodynamic limit, this $N_d^2$ scaling diverges faster than the system size, indicating an instability towards superconductivity at $T=0$.

\begin{figure*}
	\includegraphics[width=1.0\textwidth]{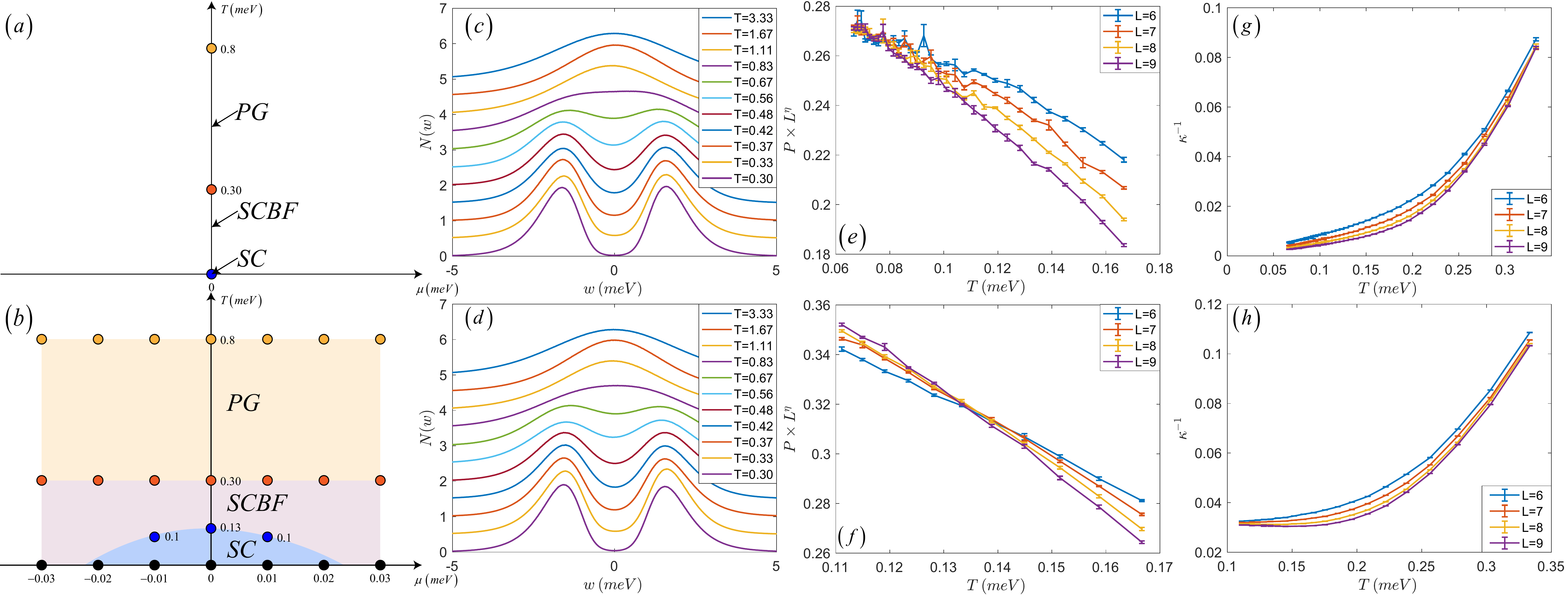}
\caption{QMC simulations of the flat-band limit (top row) and away from the flat-band limit (bottom row). The bandwidths of non-interacting band structures are set to $0$ and $0.8 meV$ respectively. (a-b) Phase diagrams. In the flat-band limit (a), a finite chemical potential drives the system into a trivial insulator with either empty ($\mu\ll 0$) or complete filled ($\mu \gg 0$) bands. In Fig.~(b), the yellow/purple region is the pseudogap (PG)/ super-compressible bosonic fluid (SCBF) phase, and the blue region marks the superconducting (SC) dome. (c-d) Density of states for a $9\times9$ system at $\mu=0$. The Cooper gap survives above the superconducting transition temperature and turns into a pseudogap above $T\sim0.3$.
(e-f) Critical scaling of $P\times L^{\eta}$ versus $T$ with BKT anomalous dimension $\eta=1/4$. The crossing point of different system sizes ($L=6,7,8,9$) marks the superconducting transition temperature: $T_c\sim0$ in (e) and $\sim 0.13$ in (f).
(g-h) Inverse of compressibility $\kappa$ for $L=6,7,8,9$ at $\mu=0$. At low temperature, diverged compressibility $\kappa$ in (g) converges in superconducting phase as shown in (h).
}
	\label{fig:fig2}
\end{figure*}

Despite the finite Cooper gap and diverging superconducting correlation function, Cooper pairs in this boson fluid don't lead to superconduct at any finite temperature. This is because the superconducting order parameter is part of a SU(2) generator. Therefore, a superconducting state would spontaneously break the SU(2) symmetry, instead of just the U(1) charge symmetry. In other words, the symmetry breaking pattern here is in the Heisenberg universality class, instead of XY. For 2D systems at finite $T$, it has long been known that thermal fluctuations will destroy any order that spontaneously breaks a SU(2) symmetry (i.e., there is no finite temperature order for Heisenberg spins in 2D). Thus, although Cooper pairs have formed at $T\sim V$,  long-range or quasi-long range phase coherence cannot be developed at any finite temperature. This conclusion is verified in QMC simulations, where we observe a fully-developed Cooper gap at finite $T$, but the phase coherence remains disordered 
even down to lowest accessible temperature [Fig.~\ref{fig:fig2}(a,c,e,g)].
In addition to preventing the formation of a finite $T$ superconducting phase, the SU(2) symmetry also offers an interesting link between superconductivity fluctuations and particle-number fluctuations.
From the SU(2) symmetry, we have $\langle \sigma_z^2\rangle=\langle \sigma_x^2 \rangle$ and thus $\langle \hat{N}_p^2 \rangle - \langle \hat{N}_p \rangle^2 = 2\langle \Delta \Delta^\dagger\rangle$.
Because superconductivity fluctuations diverge as $\propto N_d^2$ at $T=0$, particle-number fluctuations  must also diverge as $\propto N_d^2$. This scaling violates one fundamental assumption of statistical physics, the central limit theorem, which requires square fluctuations to scale linearly with system sizes. Such violation is a consequence of the infinite ground-state degeneracy.

From the fluctuation-dissipation theorem, this divergence in particle-number fluctuations implies a diverging compressibility at low $T$: $\kappa=\frac{1}{N_d} \frac{d \left\langle \hat{N}_p \right\rangle}{d \mu}=\beta\frac{\left\langle \hat{N}_p^2 \right\rangle - \left\langle \hat{N}_p \right\rangle^2}{N_d}=\frac{2 \beta}{N_d} \left\langle \Delta \Delta^\dagger\right\rangle$. When temperature is reduced, $\kappa$ increases. In the low temperature limit, because $\left\langle \Delta \Delta^\dagger\right\rangle \propto N_d^2$, $\kappa$ diverges as $\kappa \propto\beta N_d$. This divergence is also seen in QMC simulations in Fig~\ref{fig:fig2}(g).

In summary, in the flat-band limit, exact theory analysis predicts a bosonic fluid of Cooper pairs within a full Cooper gap. This bosonic fluid has a high compressibility, which diverges at $T\to 0$. To highlight this diverging compressibility, we label this state as SCBF (for super-compressible bosonic fluid) in our phase diagrams Fig.~\ref{fig:fig2}(a) and (b).

\section{ED and QMC simulations}
For simplicity, here we only consider one flat-band per valley, and choose parameters according to twisted homobilayer TMDs~\cite{HeqiuLi2021,YingSu2021} to carry out exact diagonalization (ED) and QMC simulations (see SM~\ref{app:app1}). Same techniques can also be applied to systems with more flat bands (e.g., TBGs), and all qualitative features shall remain.
We first simulate systems with sizes $3\times3$ and $3\times4$ (number of momentum points in mBZ) via ED. The number of ground states and single-particle excitations perfectly match the analytic theory
(see SM Fig.~\ref{fig:fig5}). The implementation of the momentum space QMC simulation are shown in \ref{app:app2} and \ref{app:app3} in SM, where we prove the absence of sign problem at any fillings and regardless of band width. This allows us to efficiently simulate this model (for system size up to $9\times9$) and explore the phase space both at and away from the flat-band limit.
For benchmark, we compute the single-particle Green's function and superconductor correlation function at the flat-band limit, which agree nicely with analytic theory (see SM Fig.~\ref{fig:fig4}(a,b)).

With the imaginary-time correlation functions obtained in QMC, we further employ the stochastic analytic continuation (SAC) method to extract the real-frequency spectra~\cite{PhysRevB.57.10287,beach2004identifying,PhysRevE.94.063308,PhysRevB.78.174429,PhysRevX.7.041072,CKZhou2021,GaopeiPanValley2021,sunDynamical2018,maDynamical2018,wangFractionalized2021,liKosterlitz2020,huEvidence2020}.
In Fig.~\ref{fig:fig2}(c-d), we plot the fermion DOS at different temperature for a $9\times9$ system. At $T<0.3$, a full gap is observed, which is the Cooper gap discussed above.
For $0.3<T<0.8$, fermion spectral weight start to emerge inside the gap, i.e., a pseudogap (PG) is formed.
Defining $P=\frac{1}{2 N_d^2}\left\langle \Delta \Delta^\dagger + \Delta^\dagger\Delta \right\rangle$, we also try to determine the superconducting phase transition temperature $T_c$ by probing the onset of quasi-long range order. This is achieved via data cross of $P\times L^{\eta}$ versus $T$ as shown in Fig.~\ref{fig:fig2}(e,f) and by comparing the slope of $\frac{d \ln(P)}{d \ln(L^2)}$ versus $\beta$ (see SM Fig.~\ref{fig:fig4}(c,d)) with the BKT anomalous dimension exponent $\eta=\frac{1}{4}$~\cite{Isakov2003,WLJiang2021,Pavia2004,ChuangChen2021}, using $P=L^{-\eta}f(L\cdot\exp(-\frac{A}{(T-T_c)^{1/2}}))$. In Fig.~\ref{fig:fig2}(e) for the case of flat-band limit, the cross point is indeed approaching $T_c \to 0$, confirming the absence of finite temperature phase transition, in full agreement with the exact solution.

With the exactly solvable limit understood, a kinetic energy term with finite (but small) bandwidth is introduced, which explicitly breaks the SU(2) symmetry. Here, again, we use the kinetic term of a homobilayer TMD and expands the band width to 0.8 meV. Same qualitative features are expected for other more complicated setups, such as TBGs.
Without the SU(2) symmetry,  a BKT superconducting phase becomes allowed, and in QMC simulations, we indeed observe a superconducting dome with chemical potential partially filling flat bands, in analogy to experiments reported in TBGs. Above the superconducting dome, the pseudogap and SCBF phases remain. Because the band width is still smaller than interaction energy scale, the temperature scales for the pseudogap and SCBF phases, which are dominated by interactions, are almost invariant for different band fillings (see Fig.~\ref{fig:fig2}(b) and SM Fig.~\ref{fig:fig6}), consistent with the STM experiment in TBG between different integer fillings~\cite{kerelsky2019maximized}.
Another observation in TBG experiment is $d\mu/dn$ reduces towards 0 at superconducting dopings~\cite{andrei2020graphene}, which implies a large compressibility $\kappa$. This is also seen in our simulation results.
 
 
\section{Discussion} We proposed a model describing 2D flat-band inter-valley superconductor. The exact solution and QMC simulations reveal nontrivial phenomena, such as doping independent gap and large compressibility above the  
superconducting dome, which seems consistent with experimental studies. The super-compressible fluid phase and pseudogap phases acquire intriguing features. In transport measurements, these states are conductors, but in tunneling 
experiments, they behaves like an insulator, with a finite gap/pseudogap. However, as the system is cool down to the superconducting phase, this gap evolves adiabatically across the superconducting transition, in direct contrast to an 
insulating-superconductor transition. Upon gating, the large compressibility will lead to large response in fermion density, which is a unique feature due to moir\'e flat bands  and distinguishes this bosonic fluid from other failed superconductors of non-flat bands~\cite{Kapitulnik2019}.
It is also important to point out that despite the absence of superconductivity, the boson fluid phase may exhibit certain properties of a superconductor, e.g., Andreev reflection, because all fermions have been paired up. These Cooper pairs may 
also lead to other nontrivial phenomena. For example, because charge carriers now have charge 2e, an extra factor of two may emerge in interferometry via the Aharonov-Bohm effect. The bosonic nature of the Cooper pairs may also lead to non-Fermi liquid behavior, such as the violation of Wiedemann-Franz law, absence or suppression of quantum oscillations and/or Friedel oscillations, and the departure of $C\propto T$ scaling in heat capacity.


\begin{acknowledgments}
We thank Dong-Keun Ki and Ning Wang for stimulating discussions, and Debanjan Chowdhury for very helpful suggestions. X.Z., G.P.P. and Z.Y.M. acknowledge support from the RGC of Hong Kong SAR of China (Grant Nos. 17303019, 17301420, 17301721 and AoE/P-701/20), the Strategic Priority Research Program of the Chinese Academy of Sciences (Grant No. XDB33000000), the K. C. Wong Education Foundation (Grant No. GJTD-2020-01) and the Seed Funding "QuantumInspired explainable-AI" at the HKU-TCL Joint Research Centre for Artificial Intelligence. H.L. and K.S. acknowledge support through NSF Grant No. NSF-EFMA-1741618. We thank the Computational Initiative at the Faculty of Science and the Information Technology Services at the University of Hong Kong and the Tianhe platforms at the National Supercomputer Center in Guangzhou for their technical support and generous allocation of CPU time.
\end{acknowledgments}

\begin{widetext}
	
	\section{Supplemental Material}
	\subsection{Simulation setting}
	\label{app:app1}
	In homobilayer TMDs, after two layers are rotated by a small angle $\theta$, we can see in the moir\'e Brillouin zone (mBZ) $+\mathbf{K}$ valley for the top and bottom layers are shifted to $\mathbf{K}_{t}$ and $\mathbf{K}_{b}$ (see for example, Fig.1 in Ref.~\cite{HeqiuLi2021}). A moir\'e continuum Hamiltonian~\cite{MacDonald2019,HeqiuLi2021} for the $+\mathbf{K}$ valley, which is similar to the BM model in Ref.~\cite{Bistritzer_TBG}:
	$
	H_{+}(\mathbf{k}, \mathbf{r})=\left(\begin{array}{cc}
		-\frac{\hbar^{2}\left(\mathbf{k}-\mathbf{K}_{b}\right)^{2}}{2 m^{*}}+P_{b}(\mathbf{r}) & P_{T}(\mathbf{r}) \\
		P_{T}^{\dagger}(\mathbf{r}) & -\frac{\hbar^{2}\left(\mathbf{k}-\mathbf{K}_{t}\right)^{2}}{2 m^{*}}+P_{t}(\mathbf{r})
	\end{array}\right)
	$, where $b$ and $t$ refer to bottom and top layers. $m^{*}$ is the effective mass, and $\mathbf{k}$ is momentum measured from $+\mathbf{K}$ point. Moir\'e potential $P_{b, t, T}$ can be parameterized:
	$
	P_{T}(\mathbf{r})=w\left(1+e^{-i \mathbf{G}_{2} \cdot \mathbf{r}}+e^{-i \mathbf{G}_{3} \cdot \mathbf{r}}\right), 
	P_{l}(\mathbf{r})=2 w_{z} \sum_{j=1,3,5} \cos \left(\mathbf{G}_{j} \cdot \mathbf{r}+l \psi\right)
	$, where $l \in\{b, t\}=\{+1,-1\}$ and $\mathbf{G}_{j}$ is the moir\'e reciprocal lattice vectors with length $\left|\mathbf{G}_{j}\right|=\frac{4 \pi}{\sqrt{3} a_{M}}$ and polar angle $\frac{\pi(j-1)}{3}$. Here $a_{M}=a_{0} / \theta$ is the moir\'e lattice constant when $\theta$ is small. These  parameters have been obtained from the first-principle calculations for $MoTe_2$ homobilayer $\left(\hbar^{2} / 2 m^{*} a_{0}^{2}, w_{z}, w, \psi\right)=$ $\left(495 \mathrm{meV}, 8 \mathrm{meV},-8.5 \mathrm{meV},-89.6^{\circ}\right)$\cite{MacDonald2019}. The Hamiltonian of the other valley $-K$ can be obtained by applying the time-reversal operation: $H_{-}(\mathbf{k}, \mathbf{r})=H_{+}(-\mathbf{k}, \mathbf{r})^{*}$.
	
	Our form factor $\lambda_{m,n,\tau}(\mathbf{k}, \mathbf{k+q+G})$ is defined by $\lambda_{m,n,\tau}(\mathbf{k}, \mathbf{k+q+G}) = \sum_{\mathbf{G}^{\prime}, X} u_{m, \tau ; \mathbf{G}^{\prime}, X}^{*}(\mathbf{k}) u_{n, \tau ; \mathbf{G}^{\prime}+\mathbf{G}, X}(\mathbf{k+q})$, where $u_{m, \tau ; \mathbf{G}^{\prime}, X}^{*}(\mathbf{k})$ is unitary transiformation matrix linking plane-wave and band basis, while the index $X$ represents all other degrees of freedom, such as layer, sublattice, spin indices, etc.
	To describe twisted TMDs, one can simply discard this subindex $X$.
	
	For our interaction $V(\bq)$, we use double-gate screened Coulomb interaction
	\begin{equation}
		\frac{V(\bq)}{\Omega}\approx \frac{\theta}{N_k} \frac{4\pi}{\sqrt{3}} \frac{\tanh(\bq\cdot \bf{d})}{\bq\cdot \bf{a_M}} meV
	\end{equation}
	Here $\theta$ is twist angle $1.38^{\circ}$, $N_k$ is number of momentum points in mBZ and $\bf{d}$ is the distance between two screened gates set as $d=2a_{M}$. At this twist angle, the dispersion of the top three bands is plotted in Fig.~\ref{fig:fig3}.
	
	\begin{figure}[htp!]
		\includegraphics[width=0.6\columnwidth]{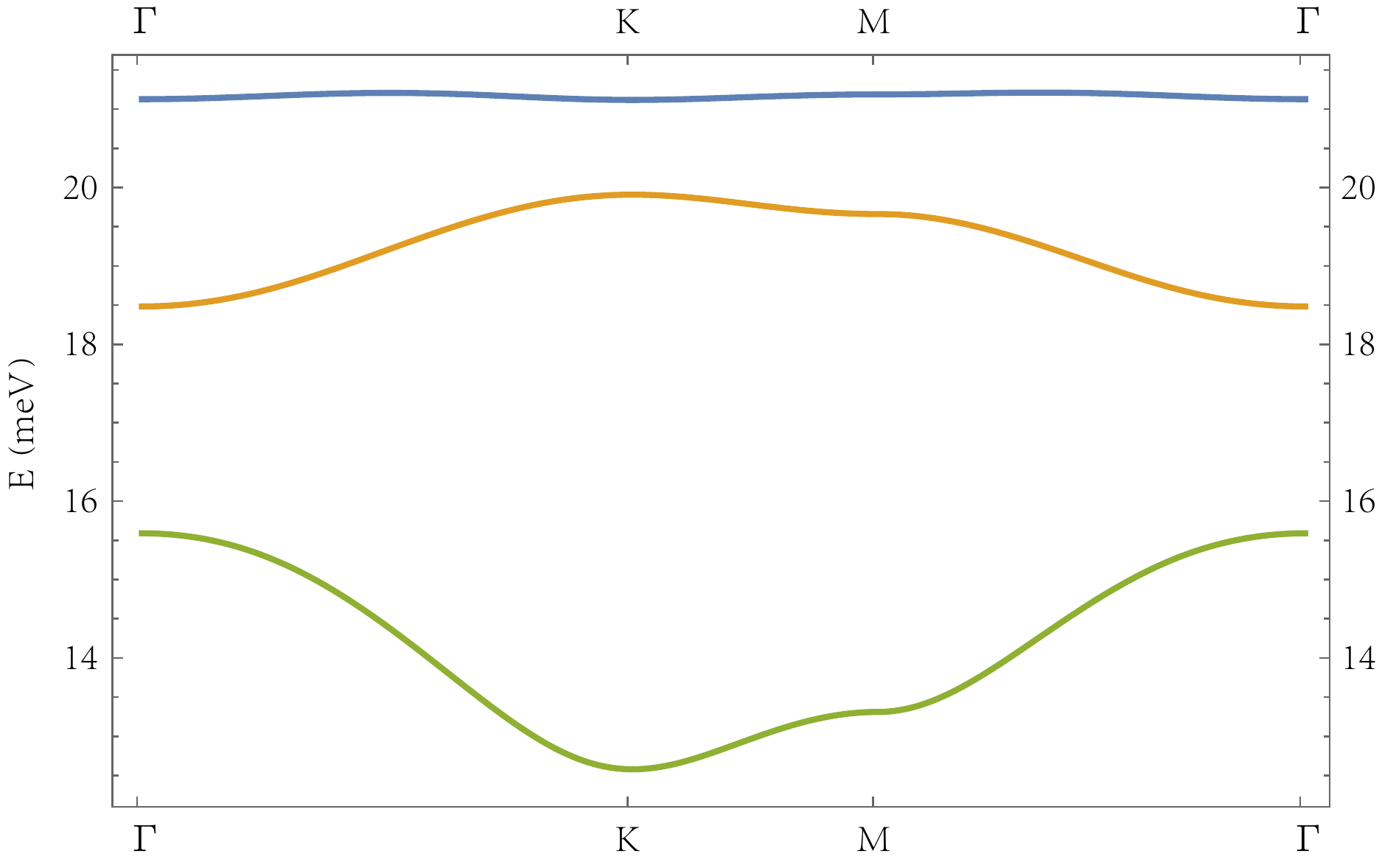}
		\caption{Dispersion of the top three bands in mBZ.}
		\label{fig:fig3}
	\end{figure}
	
	\subsection{Implementation of QMC}
	\label{app:app2}
	We follow the implementation of momentum space quantum Monte Carlo developed by us in Ref.~\cite{XuZhang2021}. Starting from Hamiltonian in flat band basis
	
	\begin{eqnarray}
		H_{I}&=&\frac{1}{2 \Omega} \sum_{\mathbf{G}} \sum_{\mathbf{q} \in mBZ} V(\mathbf{q+G}) \delta \rho_{\mathbf{q+G}} \delta \rho_{\mathbf{-q-G}} \nonumber\\
		\delta \rho_{\mathbf{q+G}}&=&\sum_{\mathbf{k},m,n} [\lambda_{m,n,\tau}(\mathbf{k}, \mathbf{k+q+G})c_{\mathbf{k}, m, \tau}^{\dagger} c_{\mathbf{k+q}, n, \tau} -\lambda_{m,n,-\tau}(\mathbf{k}, \mathbf{k+q+G})c_{\mathbf{k},m,-\tau}^{\dagger} c_{\mathbf{k+q},n,-\tau}]
	\end{eqnarray}
	
	According to the discrete Hubbard-Stratonovich transformation, $e^{\alpha \hat{O}^{2}}=\frac{1}{4} \sum_{l=\pm 1,\pm 2} \gamma(l) e^{\sqrt{\alpha} \eta(l) \hat{o}}+O\left(\alpha^{4}\right)$, 
	where $\gamma(\pm 1)=1+\frac{\sqrt{6}}{3}$, $\gamma(\pm 2)=1-\frac{\sqrt{6}}{3}$, $\eta(\pm 1)=\pm \sqrt{2(3-\sqrt{6})}$ and $\eta(\pm 2)=\pm \sqrt{2(3+\sqrt{6})}$, we can rewrite the partition function as,
	
	\begin{eqnarray}
		&Z&=\Tr\{\prod_{t}e^{-\Delta \tau H_{I}(t)}\} =\Tr\{\prod_{t} e^{-\Delta \tau \frac{1}{4 \Omega}\sum_{|\bq+\bG|\neq0} V(\bq+\bG)\left[\left(\delta\rho_{-\bq-\bG}+\delta\rho_{\bq+\bG}\right)^{2}-\left(\delta\rho_{-\bq-\bG}-\delta\rho_{\bq+\bG}\right)^{2}\right]  }\} \nonumber\\
		&\approx& \sum_{\{l_{|\bq|,t}\}} \prod_{t} [ \prod_{|\bq+\bG|\neq0}\frac{1}{16} \gamma\left(l_{|\bq|_1,t}\right) \gamma\left(l_{|\bq|_2,t}\right)]  \Tr\{\prod_{t}[\prod_{|\bq+\bG|\neq0}e^{i \eta\left(l_{|\bq|_1,t}\right) A_{\bq}\left(\delta\rho_{-\bq}+\delta\rho_{\bq}\right)} e^{\eta\left(l_{|\bq|_2,t}\right) A_{\bq}\left(\delta\rho_{-\bq}-\delta\rho_{\bq}\right)}]\}
	\end{eqnarray}
	
	Here $t$ is the imaginary time index with step $\Delta\tau$, $A_{\bq+\bG} =\sqrt{\frac{\Delta \tau}{4} \frac{V(\bq+\bG)}{\Omega}}$ and $\{l_{|\bq|_1,t},l_{|\bq|_2,t}\}$ are the four-component auxiliary fields.
	
	Generally, average of any observables $\hat{O}$ can be written as,
	
	\begin{equation}
		\langle \hat{O} \rangle = \frac{\Tr(\hat{O} e^{-\beta H})}{\Tr(e^{-\beta H})} = \sum_{\{l_{|\bq|,t}\}} \frac{P(\{l_{|\bq|,t}\}) \Tr[\prod_{t}\hat{B}_t(\{l_{|\bq|,t}\})] \frac{\Tr[\hat{O} \prod_{t}\hat{B}_t(\{l_{|\bq|,t}\})]}{\Tr[\prod_{t}\hat{B}_t(\{l_{|\bq|,t}\})]}}{\sum_{\{l_{|\bq|,t}\}} P(\{l_{|\bq|,t}\}) \Tr[\prod_{t}\hat{B}_t(\{l_{|\bq|,t}\})]} 
	\end{equation}
	
	$P(\{l_{|\bq|,t}\})=\prod_{t} [ \prod_{|\bq+\bG|\neq0}\frac{1}{16} \gamma\left(l_{|\bq|_1,t}\right) \gamma\left(l_{|\bq|_2,t}\right)]$, $\hat{B}_t(\{l_{|\bq|,t}\})=\prod_{|\bq+\bG|\neq0}e^{i \eta\left(l_{|\bq|_1,t}\right) A_{\bq}\left(\delta\rho_{-\bq}+\delta\rho_{\bq}\right)} e^{\eta\left(l_{|\bq|_2,t}\right) A_{\bq}\left(\delta\rho_{-\bq}-\delta\rho_{\bq}\right)}$, respectively. We see $P_l=P(\{l_{|\bq|,t}\}) \Tr[\prod_{t}\hat{B}_t(\{l_{|\bq|,t}\})]$ as possibility weight and $\langle \hat{O} \rangle_l=\frac{\Tr[\hat{O} \prod_{t}\hat{B}_t(\{l_{|\bq|,t}\})]}{\Tr[\prod_{t}\hat{B}_t(\{l_{|\bq|,t}\})]}$ as sample value for configuration $\{l_{|\bq|,t}\}$. Then Markov chain Mento Carlo can compute this $\langle \hat{O} \rangle$.
	
	\subsection{Absence of Sign problem}
	\label{app:app3}
	Here, we prove there is no sign problem for our Hamiltonian. Single-particle matrixes between two valleys satisfy,
	\begin{equation}
		\begin{aligned}
			\delta \rho_{\mathbf{q+G}, \tau}&=\sum_{\mathbf{k},m,n}\left[\lambda_{m,n,\tau}(\mathbf{k}, \mathbf{k+q+G})c_{\mathbf{k}, m,\tau}^{\dagger} c_{\mathbf{k+q}, n,\tau}\right] \\
			\delta \rho_{\mathbf{q+G},-\tau}&=\sum_{\mathbf{k},m,n}\left[-\lambda_{m,n,-\tau}(\mathbf{k}, \mathbf{k+q+G})c_{\mathbf{k},m,-\tau}^{\dagger} c_{\mathbf{k+q},n,-\tau}\right] \\
			&=\sum_{\mathbf{k},m,n}\left[-\lambda_{m,n,\tau}^{*}(\mathbf{k}, \mathbf{k-q-G})c_{\mathbf{-k},m,-\tau}^{\dagger} c_{\mathbf{-k+q},n,-\tau}\right] \\
			&=\sum_{\mathbf{k},m,n}\left[-\lambda_{m,n,\tau}^{*}(\mathbf{k}, \mathbf{k-q-G})\tilde{c}_{\mathbf{k},m,-\tau}^{\dagger} \tilde{c}_{\mathbf{k-q},n,-\tau}\right] \\
			&=-\delta \rho_{\mathbf{-q-G}, \tau}^{*}
		\end{aligned}
	\end{equation}
	Here, $\tilde{c}_{\mathbf{k},m,-\tau}=c_{\mathbf{-k},m,-\tau}$. Then one can see even with flat band kinetic terms,
	\begin{equation}
		\begin{aligned}
			\hat{B}_{t, \tau}\left(\left\{l_{|\bq|, t}\right\}\right)&=e^{-\Delta \tau H_{0, \tau}} \prod_{|\bq| \neq 0} e^{i \eta\left(l_{\left|\bq_{1}\right|,t}\right) A_{\bq}\left(\delta \rho_{-\bq, \tau}+\delta \rho_{\bq, \tau}\right)} e^{\eta\left(l_{\left|\bq_{2}\right|, t}\right) A_{\bq}\left(\delta \rho_{-\bq, \tau}-\delta \rho_{\bq, \tau}\right)} \\
			\hat{B}_{t,-\tau}\left(\left\{l_{|\bq|, t}\right\}\right)&=e^{-\Delta \tau H_{0,-\tau}} \prod_{|\bq| \neq 0} e^{i \eta\left(l_{|\bq_1|, t}\right) A_{\bq}\left(-\delta \rho_{\bq,\tau}^{*}-\delta \rho_{-\bq,\tau}^{*}\right)} e^{\eta\left(l_{|\bq_2|, t}\right) A_{\bq}\left(-\delta \rho_{\bq,\tau}^{*}+\delta \rho_{-\bq,\tau}^{*}\right)}=\hat{B}_{t, \tau}^{*}\left(\left\{l_{|\bq|, t}\right\}\right) \\
			\operatorname{Tr}\left[\prod_{t} \hat{B}_{t}\left(\left\{l_{|\bq|, t}\right\}\right)\right]&=\operatorname{Tr}\left[\prod_{t} \hat{B}_{t, \tau}\left(\left\{l_{|\bq|, t}\right\}\right)\right] \cdot \operatorname{Tr}\left[\prod_{t} \hat{B}_{t,-\tau}\left(\left\{l_{|\bq|, t}\right\}\right)\right]=\left|\operatorname{Tr}\left[\prod_{t} \hat{B}_{t, \tau}\left(\left\{l_{|\bq|, t}\right\}\right)\right]\right|^{2}
		\end{aligned}
	\end{equation}
	This is always a non-negative number so that there is no sign problem.
	
	\subsection{Details for exact solution}
	\label{app:app4}
	First, we show the $N_d+1$ degenerate ground states belong to one $N_d+1$ dimensional irrep of SU(2), which can be represented by normal Young diagram below
	\begin{center}
		$\underbrace{\ydiagram{4}\cdots\cdots\ydiagram{4}}$ \\
		$N_d$
	\end{center}
	The dimension of irrep can be calculated by hook's rule
	\begin{equation}
		d_{\left[ N_d\right] }(SU(2)) = \prod_{j}\frac{2+j-1}{j} = \frac{(N_d+1)!}{N_d!} = N_d+1
	\end{equation}
	
	Then, we derive single-particle excitation of Hamiltonian in Eq.~\eqref{eq:eq1}. In the single-particle Hilbert subspace, $H_I = \frac{1}{2 \Omega} \sum_{\mathbf{G},\mathbf{q}} V(\mathbf{q+G}) \sum_{\mathbf{k},m,n,n^{\prime},\tau} \lambda_{m,n^{\prime},\tau}(\mathbf{k}, \mathbf{k+q+G}) \lambda_{n^{\prime},n,\tau}(\mathbf{k+q+G}, \mathbf{k})c_{\mathbf{k}, m, \tau}^{\dagger} c_{\mathbf{k}, n, \tau}$, where $m,n,n'$ are flat band labels. One can see there is no inter-valley term, so it is expected this single-particle spectrum is the same as that in the inter-valley repulsion Hamiltonian, which can be seen the red lines have same dispersion as shown in Fig.~\ref{fig:fig1}. By diagonalizing $[\lambda_{\tau}(\mathbf{k}, \mathbf{k+q+G})\lambda_{\tau}^{\dagger}(\mathbf{k}, \mathbf{k+q+G})]_{m,n}$, we can obtain excited eigenstates $\left| \psi_{1,\mathbf{k},\tau}\right\rangle = c_{\mathbf{k},\tau}^{\dagger}\left| 0\right\rangle$. Then $(\Delta^{\dagger})^n c_{\mathbf{k},\tau}^{\dagger} \left| 0\right\rangle$ is also an excited eigenstate with the same eigenvalue $\varepsilon_{\mathbf{k},\tau}$.
	
	We would like to normalize the ground state and single-particle excitation states above it, such as,
	$\left| \psi_{2n}\right\rangle = \sqrt{\frac{(N_d-n)!}{n!N_d!}}(\Delta^{\dagger})^n\left| 0\right\rangle$, $\left| \psi_{2n+1,\mathbf{k},\tau}\right\rangle = \sqrt{\frac{(N_d-n-1)!}{n!(N_d-1)!}}(\Delta^{\dagger})^n c_{\mathbf{k},\tau}^{\dagger}\left| 0\right\rangle$, and $c_{\mathbf{k},\tau}^{\dagger} \left| \psi_{2n}\right\rangle = \sqrt{\frac{N_d-n}{N_d}} \left| \psi_{2n+1,\mathbf{k},\tau}\right\rangle$, $c_{\mathbf{k},\tau} \left| \psi_{2n}\right\rangle = \sqrt{\frac{n}{N_d}} \left| \psi_{2n-1,\mathbf{-k},-\tau}\right\rangle$, and $\Delta^\dagger \left| \psi_{2n}\right\rangle = \sqrt{(N_d-n)(n+1)} \left| \psi_{2n+2}\right\rangle$. Here $\left| \psi_{2n}\right\rangle$ and $\left| \psi_{2n\pm1,\mathbf{k},\tau}\right\rangle$ are normalized eigenstates with $2n$ and $2n\pm1$ electrons. According to those normalization relations, we can derive single-particle Green's function at zero temperature limit,
	\begin{eqnarray}
		G_{\mathbf{k},\tau}(t) &=& \frac{\Tr(e^{-(\beta-t)H_I}c_{\mathbf{k},\tau}e^{-t H_I}c_{\mathbf{k},\tau}^{\dagger})}{\Tr(e^{-\beta H_I})} \nonumber\\
		&\stackrel{\lim\limits_{\beta\to\infty}}{=}& \frac{1}{N_d+1} [ \sum_{n=0}^{N_d-1} e^{-t \varepsilon_{\mathbf{k},\tau}}\left| \left\langle c_{\mathbf{k},\tau}^{\dagger} \psi_{2n} |  \psi_{2n+1,\mathbf{k},\tau}\right\rangle \right| ^2 \nonumber\\
		&& + \sum_{n=1}^{N_d} e^{-(\beta-t) \varepsilon_{\mathbf{k},\tau}}\left| \left\langle \psi_{2n-1,-\mathbf{k},-\tau} | c_{\mathbf{k},\tau} \psi_{2n}\right\rangle \right| ^2 ] \nonumber\\
		&=& \frac{1}{2} \left[ e^{-t \varepsilon_{\mathbf{k},\tau}} + e^{-(\beta-t) \varepsilon_{\mathbf{k},\tau}}\right],
		\label{eq:eq5}
	\end{eqnarray}
	note we use $t\in[0,\beta]$ instead of the usual $\tau$  to represent the imaginary time as $\tau$ has been occupied as valley index. 
	Besides, we can also exactly derive imaginary time correlation of Cooper pair operators at zero temperature.
	\begin{eqnarray}
		\left\langle \Delta(t) \Delta^\dagger(0)\right\rangle  &=& \frac{\Tr(e^{-(\beta-t)H_I} \Delta e^{-t H_I}\Delta^\dagger)}{\Tr(e^{-\beta H_I})} \nonumber\\
		&\stackrel{\lim\limits_{\beta\to\infty}}{=}& \frac{1}{N_d+1} \sum_{n=0}^{N_d-1} \left| \left\langle \Delta^\dagger \psi_{2n} | \psi_{2n+2} \right\rangle \right|^2 \nonumber\\
		&=& \frac{N_d(N_d+2)}{6}.
		\label{eq:eq6}
	\end{eqnarray}
	Since pairing correlation function $P$ is defined as $P=\frac{1}{2 N_d^2}\left\langle \Delta \Delta^\dagger + \Delta^\dagger\Delta \right\rangle$, we actually achieve $P$ at zero temperature.
	
	Next, following the proof of statement 4 in Ref.~\cite{XuZhang2021}, we formulate in QMC framework and give a proof of relation $\langle \hat{N}_p^2 \rangle - \langle \hat{N}_p \rangle^2 = 2 \langle \Delta \Delta^\dagger \rangle$ when there is no kinetic term in our Hamiltonian.
	
	One can see $\hat{B}_{t, \tau}\left(\left\{l_{|\bq|, t}\right\}\right)$ is an unitary operator for any configuration $\left\{l_{|\bq|, t}\right\rbrace $. In single-particle basis, we write the matrix form of $\prod_{t}\hat{B}_{t, \tau}\left(\left\{l_{|\bq|, t}\right\}\right)$ as $U=e^{M_1}e^{M_2}\cdots e^{M_n}$. According to QMC's formula, Green's function for this configuration is defined as
	
	\begin{equation}
		G_{i,j}(\tau)=\frac{\operatorname{Tr}\left[c_{i,\tau} c_{j,\tau}^\dagger \prod_{t} \hat{B}_{t,\tau}\left(\left\{l_{|\bq|, t}\right\}\right) \right]}{\operatorname{Tr}\left[\prod_{t} \hat{B}_{t,\tau}\left(\left\{l_{|\bq|, t}\right\}\right)\right]}=[(I+U)^{-1}]_{i,j}
	\end{equation}
	
	By seeing $G(\tau)+G^{\dagger}(\tau)=(I+U)^{-1}+(I+U^{-1})^{-1}=(I+U)^{-1}+U(I+U)^{-1}=I$, we have $G_{i,j}(\tau)+G_{j,i}^*(\tau)=\delta_{i,j}$. To compute particle fluctuations, we can write $\langle \hat{N}_p^2 \rangle_l $ and $\langle \Delta \Delta^\dagger \rangle_l $ as
	
	\begin{eqnarray}
		\langle \Delta \Delta^\dagger \rangle_l &=& \langle \sum_{\mathbf{k}_1,m_1} c_{-\mathbf{k}_1,m_1,-\tau} c_{\mathbf{k}_1,m_1,\tau} \sum_{\mathbf{k}_2,m_2} c_{\mathbf{k}_2,m_2,\tau}^{\dagger} c_{-\mathbf{k}_2,m_2,-\tau}^{\dagger} \rangle_l \nonumber\\
		&=& \sum_{\mathbf{k}_1,m_1} \sum_{\mathbf{k}_2,m_2} \langle c_{-\mathbf{k}_1,m_1,-\tau} c_{-\mathbf{k}_2,m_2,-\tau}^{\dagger} \rangle_l \langle c_{\mathbf{k}_1,m_1,\tau} c_{\mathbf{k}_2,m_2,\tau}^{\dagger} \rangle_l \nonumber\\
		&=& \sum_{\mathbf{k}_1,m_1} \sum_{\mathbf{k}_2,m_2} \left| G_{\mathbf{k}_1 m_1,\mathbf{k}_2 m_2}\right|^2 \nonumber\\
		\langle \hat{N}_p^2 \rangle_l &=& \langle \sum_{\mathbf{k}_1,m_1,\tau_1} c_{\mathbf{k}_1,m_1,\tau_1}^\dagger c_{\mathbf{k}_1,m_1,\tau_1} \sum_{\mathbf{k}_2,m_2,\tau_2} c_{\mathbf{k}_2,m_2,\tau_2}^\dagger c_{\mathbf{k}_2,m_2,\tau_2} \rangle_l \nonumber\\
		&=& \sum_{\mathbf{k}_1,m_1,\tau_1} \sum_{\mathbf{k}_2,m_2,\tau_2} \langle c_{\mathbf{k}_1,m_1,\tau_1}^\dagger c_{\mathbf{k}_1,m_1,\tau_1} \rangle_l \langle c_{\mathbf{k}_2,m_2,\tau_2}^\dagger c_{\mathbf{k}_2,m_2,\tau_2} \rangle_l + \langle c_{\mathbf{k}_1,m_1,\tau_1}^\dagger c_{\mathbf{k}_2,m_2,\tau_2} \rangle_l \langle c_{\mathbf{k}_1,m_1,\tau_1} c_{\mathbf{k}_2,m_2,\tau_2}^\dagger \rangle_l \nonumber\\
		&=& \left(\sum_{\mathbf{k}_1,m_1} 2 - \left[ G_{\mathbf{k}_1,m_1}(\tau)+G_{\mathbf{k}_1,m_1}^*(\tau)\right]  \right)^2 + \sum_{\mathbf{k}_1,m_1} \sum_{\mathbf{k}_2,m_2} \sum_{\tau} \left[ \delta_{\mathbf{k}_1,\mathbf{k}_2}\delta_{m_1,m_2}-G_{\mathbf{k}_2 m_2,\mathbf{k}_1 m_1}(\tau)\right] G_{\mathbf{k}_1 m_1,\mathbf{k}_2 m_2}(\tau) \nonumber\\
		&=& N_d^2 + 2\sum_{\mathbf{k}_1,m_1} \sum_{\mathbf{k_2},m_2} \left| G_{\mathbf{k}_1 m_1,\mathbf{k}_2 m_2}\right|^2 \nonumber\\
		&=& N_d^2 + 2\left\langle \Delta \Delta^\dagger \right\rangle_l
	\end{eqnarray}
	
	Since we can also easily see $\langle \hat{N}_p \rangle_l = N_d$, after averaging all configurations, one will get $\langle \hat{N}_p^2 \rangle - \langle \hat{N}_p \rangle^2 = 2\langle \Delta \Delta^\dagger\rangle$.
	
	Finally, we would like to derive two-fermion excitations following similar method in Ref.~\cite{bernevig2020tbg5}. For $\bp\neq0$, it is easy to check $\left\langle 0\left| \Delta^n c_{-\bk_2-\bp,-\tau} c_{\bk_2,\tau}c^{\dagger}_{\bk_1,\tau} c^{\dagger}_{-\bk_1-\bp,-\tau} (\Delta^{\dagger})^n \right| 0\right\rangle = \delta_{\bk_1,\bk_2} A$ and $\left\langle 0\left| \Delta^n c^{\dagger}_{\bk_2+\bp,\tau} c_{\bk_2,\tau}c^{\dagger}_{\bk_1,\tau} c_{\bk_1+\bp,\tau} (\Delta^{\dagger})^n \right| 0\right\rangle = \delta_{\bk_1,\bk_2} A$ where $A$ is a normalization constant. This means two-fermion excitations on ground states $c^{\dagger}_{\bk_1,\tau} c^{\dagger}_{-\bk_1-\bp,-\tau} (\Delta^{\dagger})^n \left| 0\right\rangle$ or $c^{\dagger}_{\bk_1,\tau} c_{\bk_1+\bp,\tau} (\Delta^{\dagger})^n \left| 0\right\rangle$ are orthogonal so that they can be seen as well-defined basis. According to SU(2) symmetry, they should have the same excitation spectrum. By noticing $H_I$ applying on this basis forms a closed subspace,
	\begin{eqnarray}
		&&H_I c^{\dagger}_{\bk,\tau} c_{\bk+\bp,\tau} (\Delta^{\dagger})^n \left| 0\right\rangle \nonumber\\
		&=& \left[ H_I, c^{\dagger}_{\bk,\tau} c_{\bk+\bp,\tau}\right]  (\Delta^{\dagger})^n \left| 0\right\rangle \nonumber\\
		&=& \sum_{\bq+\bG\neq0} V(\bq+\bG) [\lambda_{\tau}(\bk,\bk+\bq+\bG)\lambda_{\tau}(\bk+\bq+\bG,\bk) c^{\dagger}_{\bk,\tau} c_{\bk+\bp,\tau} \nonumber\\
		&-&2\lambda_{\tau}(\bk+\bp,\bk+\bp+\bq+\bG)\lambda_{\tau}(\bk+\bq+\bG,\bk)c^{\dagger}_{\bk+\bq,\tau} c_{\bk+\bq+\bp,\tau} \nonumber\\
		&+&\lambda_{\tau}(\bk+\bp+\bq+\bG,\bk+\bp)\lambda_{\tau}(\bk+\bp,\bk+\bp+\bq+\bG) c^{\dagger}_{\bk,\tau} c_{\bk+\bp,\tau}] (\Delta^{\dagger})^n \left| 0\right\rangle
	\end{eqnarray}
	One can diagonalize this matrix in subspace to compute eigen excitation states as shown in Fig.~\ref{fig:fig1}(a). These excitations can be $c^{\dagger}_{\bk_1,\tau} c_{\bk_1+\bp,\tau}$ with zero charge or $c^{\dagger}_{\bk_1,\tau} c^{\dagger}_{-\bk_1-\bp,-\tau}$ with charge 2e. Thus within single-particle gap, there are continuous charged bosonic excitations.
	
	\subsection{Supplemental figures}
	\label{app:app5}
	
	Here, we use our exact solution results to benchmark the numerical code. As shown in Fig.~\ref{fig:fig4}(a-b), QMC simulations at low temperature match perfectly with the exact solution Eq.~\eqref{eq:eq5} and superconductivity pairing correlation function $P$ from QMC with increasing $\beta$ also matches the one computed from Eq.~\eqref{eq:eq6}. In Fig.~\ref{fig:fig4}(c-d), one can see the critical temperature determined by slope crossing matches well with Fig.~\ref{fig:fig2}(e-f) in the main text.
	
	We show our ED results here for $3 \times 4$ system at particle number $N=12$ and $N=11$ in Fig.~\ref{fig:fig5}. One can see one-charge excitations are gapped at all momentum points, and there are some excitations within the single-particle gap.
	
	QMC+SAC DOS results with kinetic term at different chemical potential are shown in Fig.~\ref{fig:fig6}. One can see small kinetic term with small chemical potential almost does not change single particle excitation. Also, the DOS figures below temperature $T=0.3$ ($\beta=3.3$) which are the low temperature supplement of main text Fig.~\ref{fig:fig2}(c-d) are shown in Fig.~\ref{fig:fig7}. One can see after full gapped, the position of peak is almost unchanged around single particle excitation energy. This can be understood intuitively that the DOS only measures single particle Green's function so that the pair excitation which is described by two particle Green's function as shown in Fig.~\ref{fig:fig4}(a) within the single particle gap can not be observed by DOS.
	
	Average particle number $\left\langle \hat{N}_p\right\rangle $ versus chemical potential $\mu$ is plotted with kinetic term in Fig.~\ref{fig:fig8}(a) and without kinetic term in Fig.~\ref{fig:fig8}(c) as supplement of Fig.~\ref{fig:fig2}(g-h). Due to the huge compressibility at low temperature, it is hard to compute $\frac{d \left\langle \hat{N}_p \right\rangle}{d \mu}$ by numerical differentiation precisely. We use particle fluctuation measured from QMC directly to derive the compressibility data in main text Fig.~\ref{fig:fig2}(g-h) by $\kappa=\beta\frac{\left\langle \hat{N}_p^2 \right\rangle - \left\langle \hat{N}_p \right\rangle^2}{N_d}$. The comparison of these two methods for different temperature is shown in Fig.~\ref{fig:fig8}(b,d).
	
	\begin{figure}[htp!]
		\includegraphics[width=0.8\columnwidth]{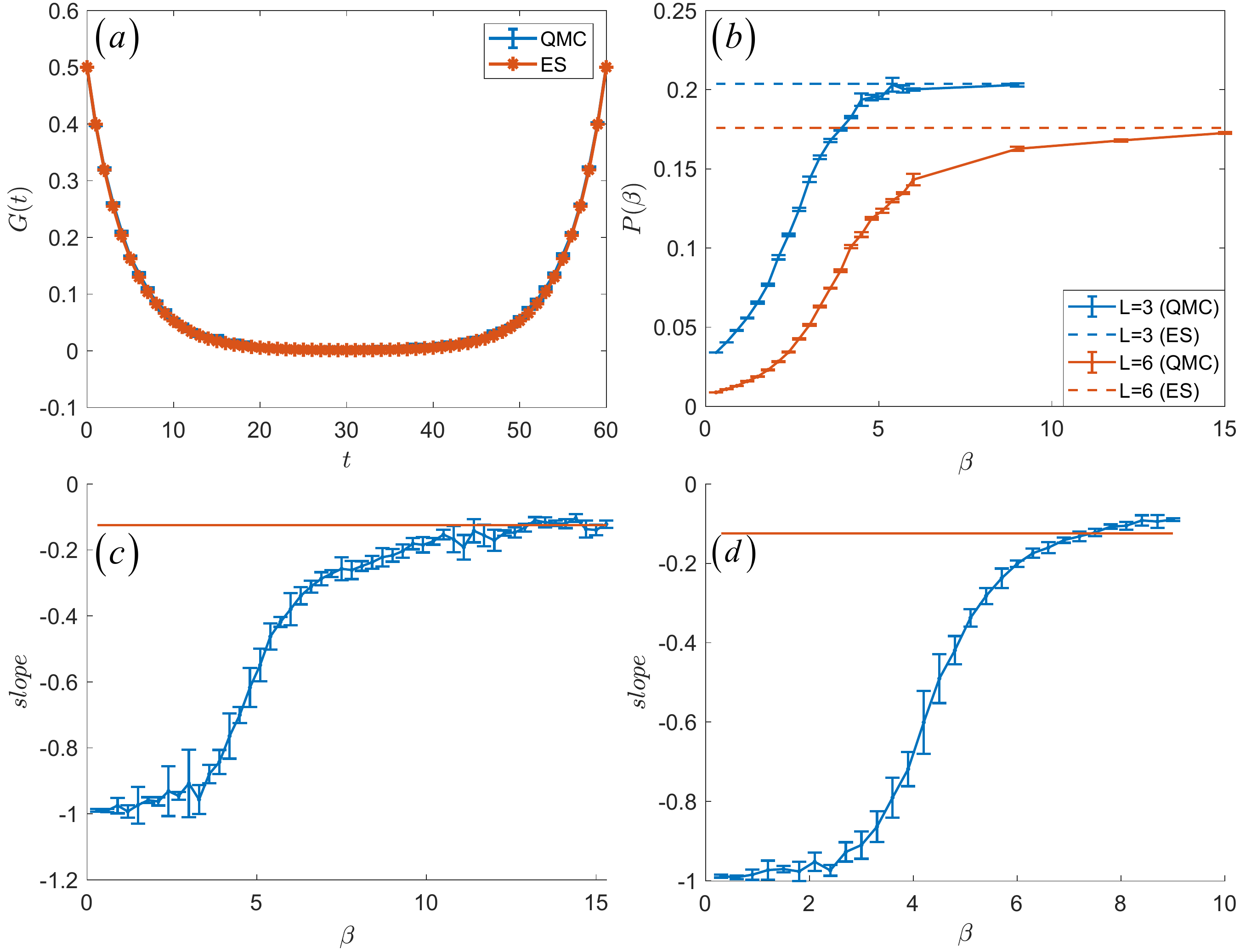}
		\caption{Benchmark results for QMC. (a-c) are results at flat band limit, (d) is simulated with kinetic term. (a) The blue line is the single-particle Green's function for $9\times9$ momentum mesh in mBZ with $\beta=6$ at $\Gamma$ point from QMC, while the red line is the exact solution (ES) according to Eq.~\eqref{eq:eq5}. (b) Superconductivity pairing correlation function for $3\times3$ and $6\times6$ systems from QMC and ES. ES result comes from Eq.~\eqref{eq:eq6}. (c-d) $slope =\frac{d \ln(P)}{d\ln(L^2)}$ vs $\beta$ at $\mu=0$. At the superconducting transition temperature, this slope shall reach $-\frac{\eta}{2}=-\frac{1}{8}$ (the red horizontal line), in good agreement with Fig.~\ref{fig:fig2}(e-f) in the main text.}
		\label{fig:fig4}
	\end{figure}
	
	\begin{figure}[htp!]
		\includegraphics[width=0.6\columnwidth]{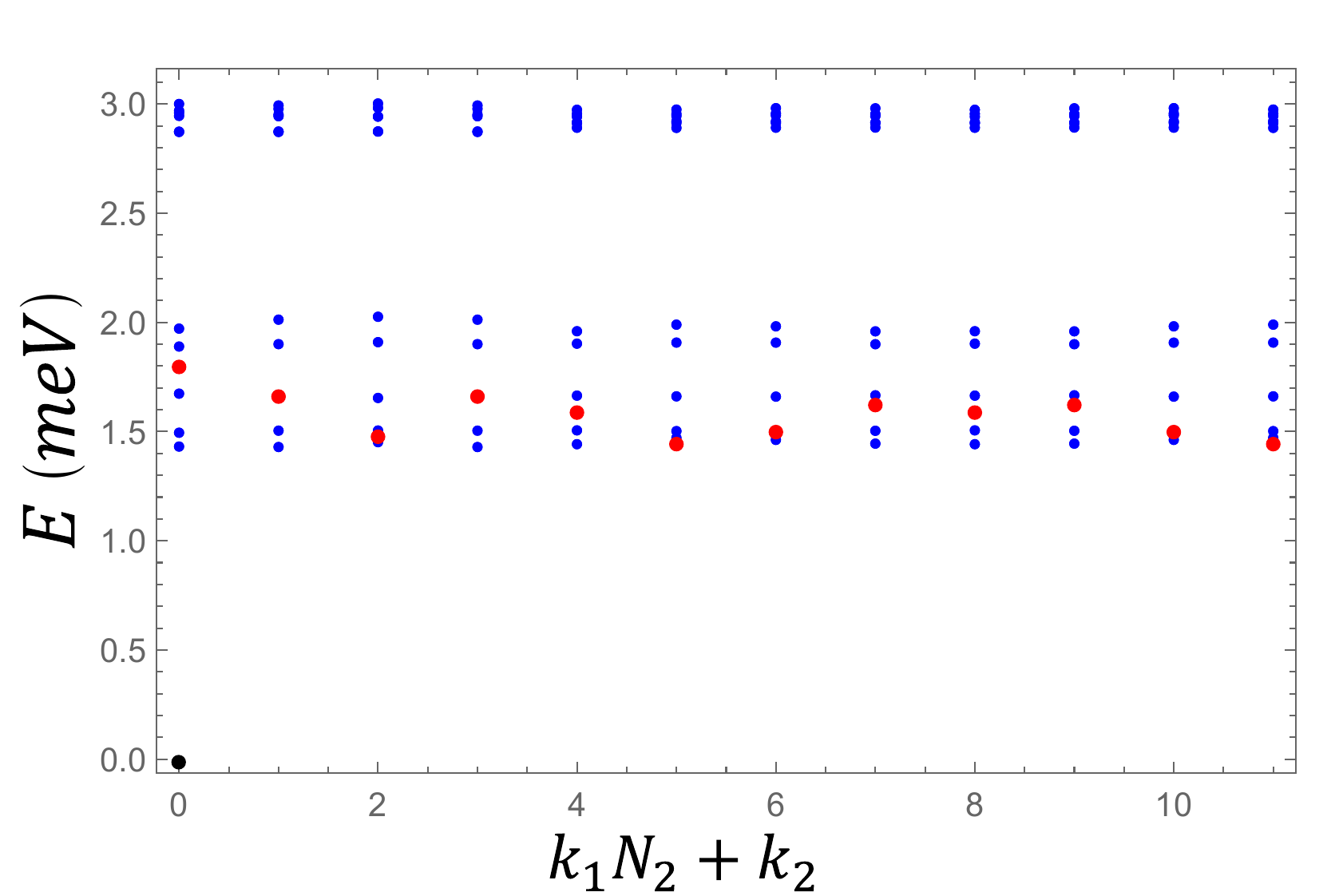}
		\caption{ED results for $3 \times 4$ system at particle number $N=12$ and $N=11$. Black point represents ground state, red points are single-particle excitations and blue points are all one-charge excitations.}
		\label{fig:fig5}
	\end{figure}
	
	\begin{figure}[htp!]
		\includegraphics[width=0.8\columnwidth]{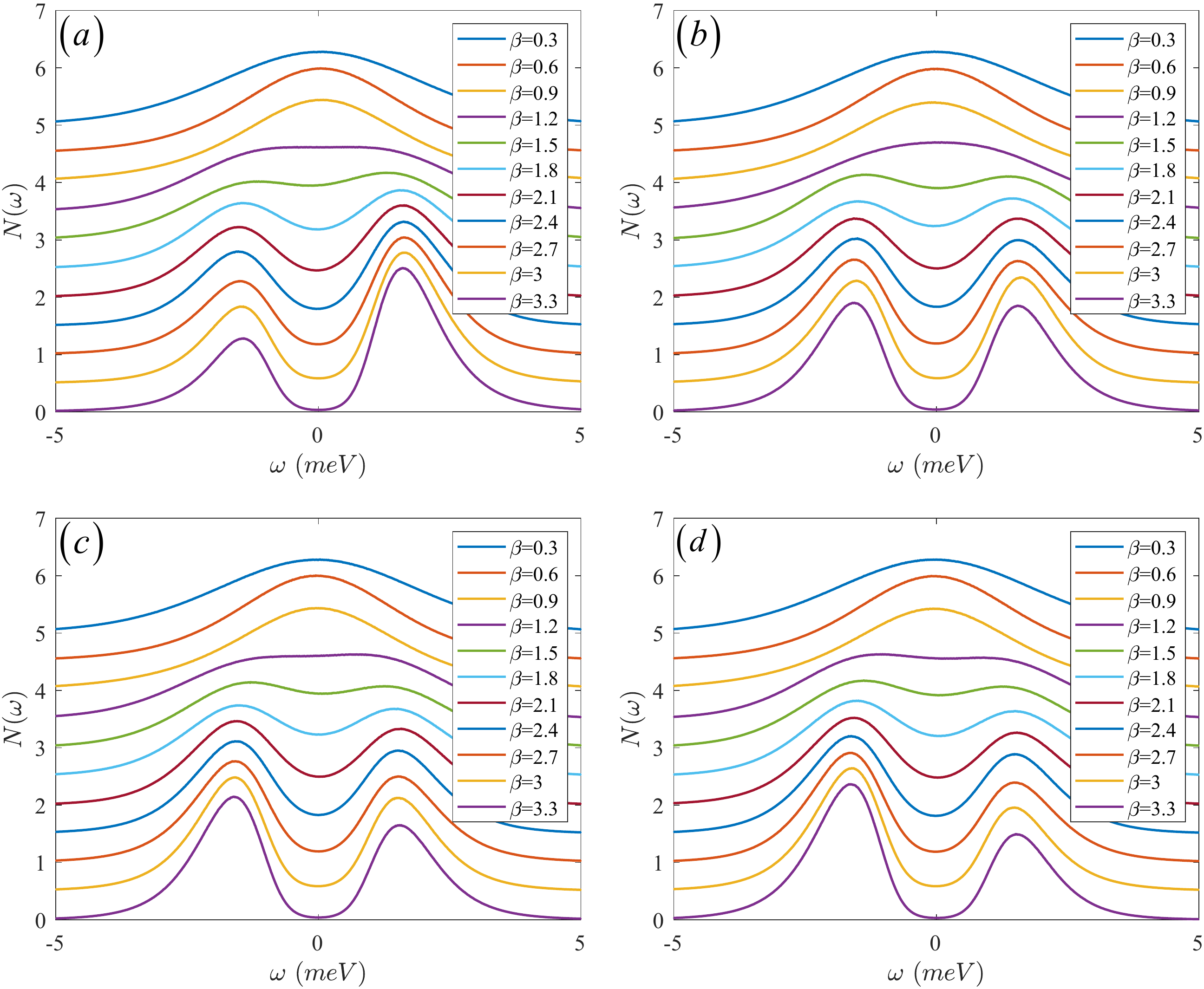}
		\caption{DOS for $9\times9$ with different $\beta$ from QMC+SAC. (a)-(d) represent $\mu=-0.03,0,0.01,0.02$ respectively.}
		\label{fig:fig6}
	\end{figure}
	
	\begin{figure}[htp!]
		\includegraphics[width=0.8\columnwidth]{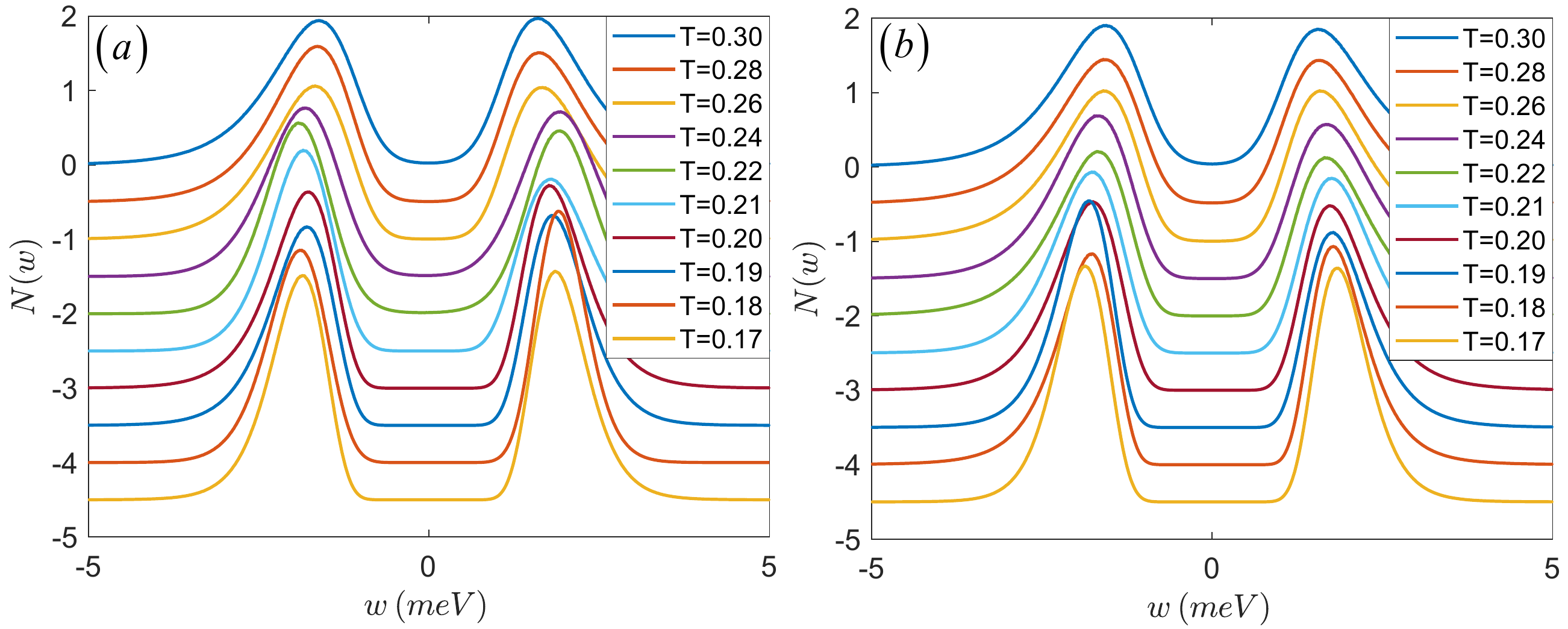}
		\caption{DOS for $9\times9$ at low temperature T from QMC+SAC. (a) DOS without kinetic term. (b) DOS with kinetic term.}
		\label{fig:fig7}
	\end{figure}
	
	\begin{figure}[htp!]
		\includegraphics[width=0.8\columnwidth]{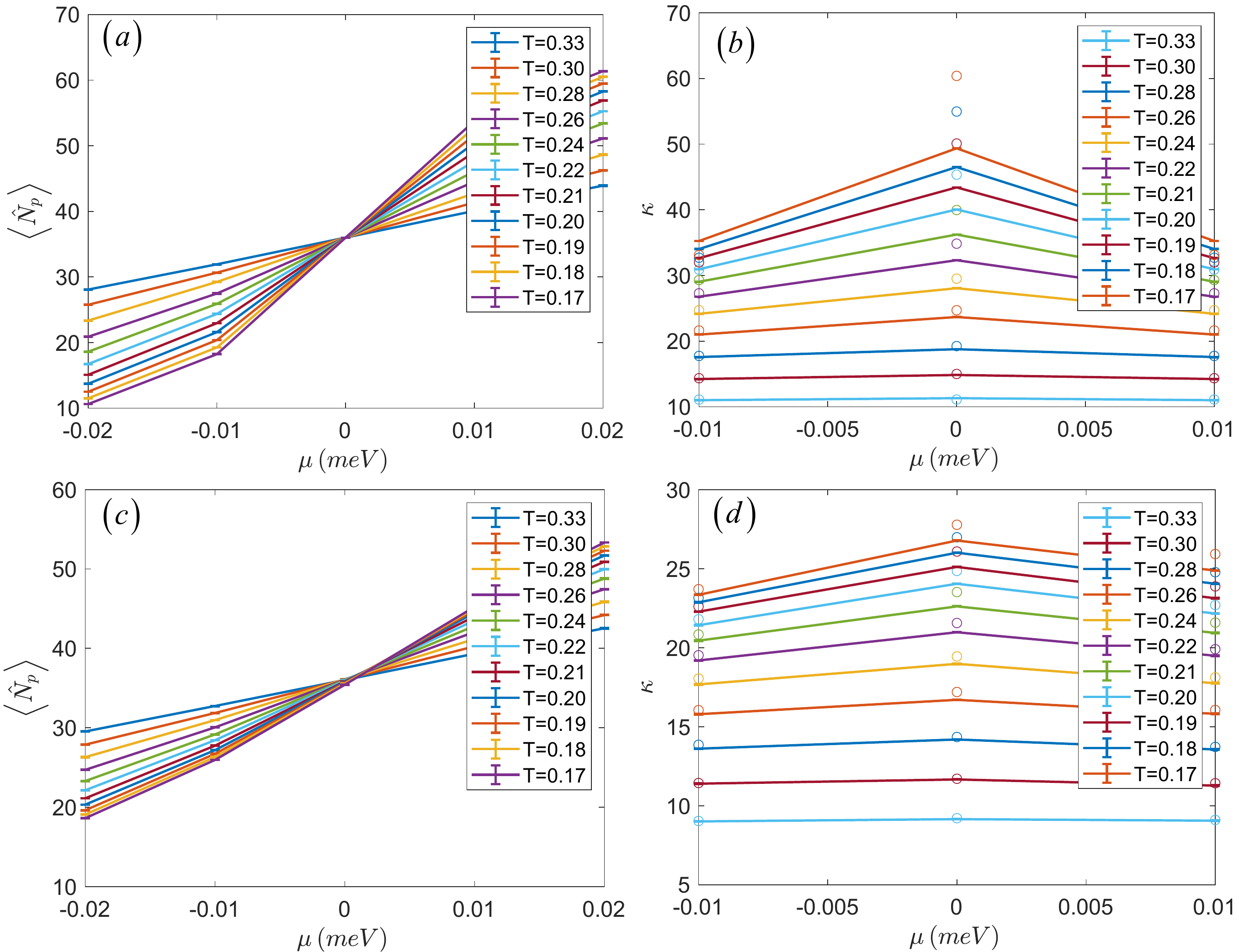}
		\caption{Average particle number $\left\langle \hat{N}_p\right\rangle $ versus chemical potential $\mu$ for $6\times6$ without kinetic term (a) and with kinetic term (c). Compressibility versus chemical potential derived from $\kappa=\frac{1}{N_d} \frac{d \left\langle \hat{N}_p \right\rangle}{d \mu}$ (colorful lines) by numerical differentiation and $\kappa=\beta\frac{\left\langle \hat{N}_p^2 \right\rangle - \left\langle \hat{N}_p \right\rangle^2}{N_d}$ (colorful circles) by QMC direct measurement without kinetic term (b) and with kinetic term (d).}
		\label{fig:fig8}
	\end{figure}
	
\end{widetext}

\clearpage

\bibliography{TMD.bib}


\end{document}